\documentclass[english,aps,prc,twocolumn,showpacs,amsmath,amssymb,preprintnumbers,
superscriptaddress]{revtex4}
\usepackage[T1]{fontenc}
\usepackage[latin9]{inputenc}
\usepackage{babel}
\usepackage{amsmath}
\usepackage{graphicx}
\usepackage{color}
\usepackage[unicode=true]{hyperref}
\usepackage{siunitx}
\usepackage{enumerate}

\newcommand{\Ga}{\Gamma}
\newcommand{\de}{\delta}

\newcommand{\ep}{\varepsilon}

\newcommand{\la}{\lambda}

\renewcommand{\th}{\theta}   % LaTeX: \th already defined

\newcommand{\om}{\omega}

\newcommand{\beq}{\begin{equation}}
\newcommand{\eeq}{\end{equation}}
\newcommand{\ba}{\begin{array}}
\newcommand{\ea}{\end{array}}
\newcommand{\bea}{\begin{eqnarray}}
\newcommand{\eea}{\end{eqnarray}}
\newcommand{\bi}{\begin{itemize}}  %\setlength{\itemsep}{0\parsep}}
\newcommand{\ei}{\end{itemize}}
\newcommand{\ben}{\begin{enumerate}} %\setlength{\itemsep}{0\parsep}}
\newcommand{\een}{\end{enumerate}}
\newcommand{\bc}{\begin{center}}
\newcommand{\ec}{\end{center}}

\newcommand{\eqn}[1]{(\ref{#1})}

\newcommand{\MeV}{{\rm MeV}}
\newcommand{\pcent}{P_{\rm cent}}
\newcommand{\Msolar}{M_\odot}

\begin{document}

\title{On the stability of strange dwarf hybrid stars}

\author{Mark G. Alford}
\affiliation{Physics Department, Washington University, St.~Louis, MO~63130, USA}
\author{Steven P. Harris}
\affiliation{Physics Department, Washington University, St.~Louis, MO~63130, USA}
\author{Pratik S. Sachdeva}
\affiliation{Department of Physics,
% Redwood Center for Theoretical Neuroscience
University of California, Berkeley, CA 94720, USA}

\begin{abstract}
We investigate the stability of stars with a density discontinuity between
a high-density core and a very low density mantle. Previous work on 
``strange dwarfs'' suggested that such a discontinuity could stabilize stars
that would have been classified as
unstable by the conventional criteria based on
extrema in the mass-radius relation. We investigate the stability of such stars by numerically
solving the Sturm-Liouville equations for the lowest-energy modes of the
star.
We find that the conventional criteria are correct, and strange dwarfs
are not stable.
\end{abstract}

\date{27 May 2017} % Hardwire date so arXiv doesn't change it

\maketitle

%%%%%%%%%%%%%%%%%%%%%%%%%%%%%%%%%%%%%%%

\section{Introduction}
\label{sec:intro}

The density profile of
stars whose temperature is low compared to the Fermi energy of their constituent
fermions can be understood in terms of a zero-temperature equation of state
$\ep(P)$ which gives the energy density as a function of the pressure.
Spherically symmetric configurations can then be obtained by solving the
Tolman-Oppenheimer-Volkoff (TOV) equation \cite{OV,Tolman}, but this procedure
yields both stable and unstable solutions.
As we will describe in more detail below, there are two commonly used methods
for determining stability of solutions to the TOV equation. One is to explicitly
obtain the eigenmodes of radial oscillation by solving 
the relevant Sturm-Liouville equation. The other, developed by
Bardeen, Thorne, and Meltzer (BTM) \cite{BTM_methods}, is based on
counting the extrema of the mass-radius relation.

It has generally been accepted that BTM proved the validity
of their criterion\cite{HHTW, thorne}. However, Glendenning and Weber have questioned this
\cite{GKW_PRL,GKW_long_paper}.  They proposed a class of configurations,
``strange dwarfs,'' that arose from an equation of state 
with a phase transition
at a relatively low critical pressure, around neutron drip.
The phase transition introduces a
large density
discontinuity between an ultra-dense strange
matter phase \cite{Bodmer:1971we,Witten:1984rs,Farhi:1984qu} and a 
much lower
density phase of conventional white-dwarf material (a degenerate electron
plasma).  Glendenning and Weber suggested that this discontinuity invalidated
the BTM mass-radius stability criteria, allowing for a family of
strange dwarf stars (with a strange matter core and a white-dwarf
mantle) that, although
occurring in a segment of the mass-radius relation that would be unstable
according to the BTM mass-radius criteria, were in fact stable.

In this paper, we investigate Glendenning and Weber's suggestion by solving
the Sturm-Liouville equation for the radial oscillation modes of
stars with a regulated discontinuity
in their equation of state. By studying the behavior of the modes as
the regulating width tends to zero, we confirm that the BTM mass-radius criteria
correctly reflect the behavior of the radial eigenmodes, and give an accurate
account of the stability of the star. This means that strange dwarfs
are not stable.

We work in natural units, where $\hbar=c=1$.

%%%%%%%%%%%%%%%%%%%%%%%%%%%%%%%%%%%%%%%%%%%%%%

\section{Two methods for determining the stability of stars}
\label{sec:methods}

There are two standard methods for determining whether a solution
of the TOV equation represents a stable star:
(a) the BTM criteria based on extrema in the mass-radius curves,
and (b) analysis of the lowest eigenmodes of radial 
oscillation \cite{BTM_methods}. 
We now describe both methods.

%%%%%%%%%%%%%%%%%%%%%%%%%%%%%%%%%%%%%%%%%%%%%%%%%%%

\subsection{The mass-radius stability criteria}
\label{sec:BTM}

For material with zero-temperature equation of state $\ep(P)$, the 
spherically symmetric gravitationally bound configurations
are described by the TOV equation,
\beq
\ba{rcl}
\dfrac{\mathop{dP}}{\mathop{dr}}
 &=&  -G \dfrac{\left(P(r)+\varepsilon(r)\right)\left(m(r)
  + 4\pi r^3 P(r)\right)}{r\left(r-2Gm(r)\right)} \ , \\[2ex]
\dfrac{\mathop{dm}}{\mathop{dr}} &=& 4\pi r^2 \varepsilon(r) \ .
\ea
\label{eq:TOV}
\eeq
where $P(r)$ and $\ep(r)$ are the pressure profile and energy-density
profile of the star, and $m(r)$ is the mass enclosed within radius $r$.
The boundary conditions are $m(0)=0$ and $P(0)$ is some chosen
central pressure $\pcent$. Integrating the TOV equation from $r=0$
outwards,
the pressure drops monotonically until at $r=R$, the radius of the star, 
the pressure reaches zero. The mass of the star is $M=m(R)$. 

The metric of the spherically symmetric, static spacetime inside the star can be written as
\beq
\mathop{ds^2} = e^{2\nu(r)}\mathop{dt^2} - e^{2\la(r)}\mathop{dr^2} - r^2\left( \mathop{d\th^2} + \sin^2{\th}\mathop{d\phi^2}  \right),
\eeq
where 
\beq
e^{2\lambda(r)}=\left(1-\frac{2Gm(r)}{r}\right)^{-1}.
\eeq
The metric coefficient $\nu(r)$ is given by \cite{glendenning_textbook}
\beq
\frac{\mathop{d\nu}}{\mathop{dr}}=\frac{G}{r}\left(\frac{ m(r)+4\pi r^3 P(r)}{r-2Gm(r)}  \right),
\label{eq:nu_ode}
\eeq
with the boundary condition $\nu(R)=(1/2)\ln(1-2GM/R)$ which ensures that it matches to the Schwarzschild solution at $r=R$. 

\begin{figure}
\includegraphics[width=\hsize]{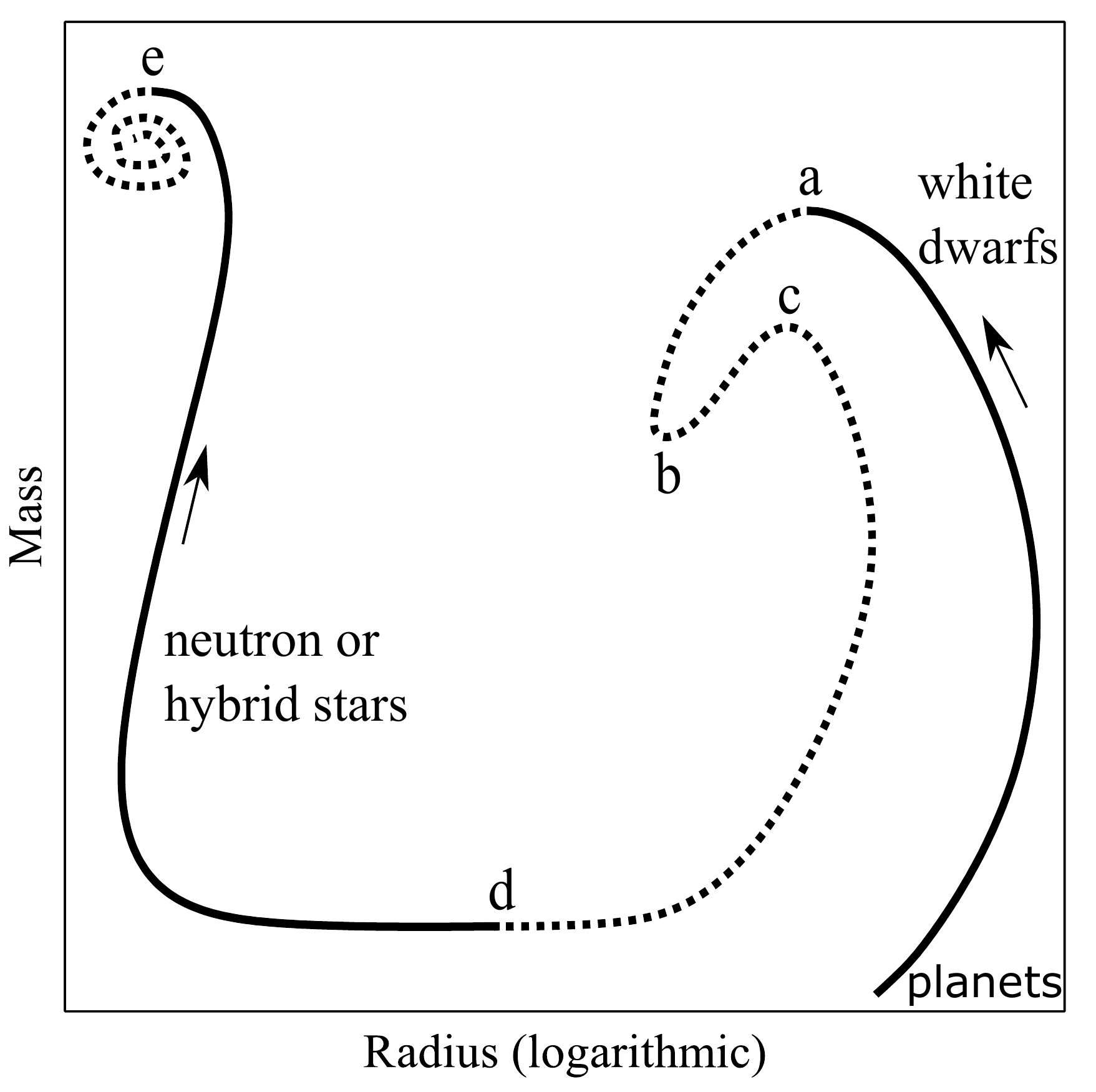}
\caption{A typical mass-radius curve of solutions to the
TOV equation for zero-temperature material \cite{hartle}.
Stable configurations are shown by solid lines, unstable configurations
by dashed lines.
At the lowest central pressure the solutions are planet-like, with
$M\propto R^3$.  As indicated by the arrows, central pressure rises as the curve goes through 5 extrema, labeled $a$ to $e$.}
\label{fig:schematic_mass_radius}
\end{figure}

By varying the central pressure
$\pcent$ one generates a one-parameter family of stationary configurations,
tracing out a curve in the mass-radius plane.
A typical curve for stars made of matter
that forms a degenerate electron gas at low pressure, and a degenerate
neutron gas at high pressure \cite{shapiro,glendenning_textbook},
 is shown in Fig.~\ref{fig:schematic_mass_radius}.
All points on the mass-radius curve are stationary configurations, but not all stationary configurations are stable against radial oscillations. 
As we will discuss in Sec.~\ref{sec:SL}, a configuration is stable
only if all its radial modes are stable.
Bardeen, Thorne, and Meltzer (BTM) \cite{BTM_methods} gave a simple 
formulation:

\beq
\parbox{0.9\hsize}{
\raggedright
BTM stability criteria:
\ben
\item At each extremum where the $M(R)$ curve rotates counter-clockwise with
increasing central pressure, one stable mode becomes unstable.
\item At each extremum where the $M(R)$ curve rotates clockwise with
increasing central pressure, one unstable mode becomes stable.
\een
}
\label{eq:BTM-criteria}
\eeq

We will now apply these criteria to Fig.~\ref{fig:schematic_mass_radius}.  The
configurations with the lowest central pressure are planet-like, with $M\ll\Msolar$
and $M\propto R^3$. These are stable \cite{thorne}. 
 As the central pressure rises, the mass
rises, giving white dwarf configurations. Then at the
Chandrasekhar mass
we reach an extremum ($a$)  where the $M(R)$ curve bends counterclockwise,
indicating that a stable mode becomes unstable.  The $M(R)$ curve then enters
an unstable interval, bending counterclockwise again through a second extremum
($b$), where a second mode becomes unstable.  At the third extremum ($c$) the
curve bends clockwise, so one of the two unstable modes becomes stable. At
the fourth extremum ($d$) the curve bends clockwise again, and the remaining
unstable mode becomes stable.  We are now on a stable branch, the ``compact
branch'', containing neutron (or hybrid) stars.  As the central pressure continues to rise the radius shrinks
rapidly and the mass rises, until we reach the fifth extremum $e$ where the
curve bends counterclockwise and the star becomes unstable. 
For typical nuclear matter equations of state, as central
pressure is increased further, the curve continues to spiral counterclockwise,
so more and more modes become unstable. 

As previously noted, Fig.~\ref{fig:schematic_mass_radius} is a typical 
mass-radius curve for stars made of degenerate electron or neutron matter.
One can explore more unusual forms of matter where this typical form of matter
has, at some critical pressure, a phase transition to an exotic phase such as quark matter.  If this critical pressure is large, then instead of spiraling at high central pressure, the mass-radius curve may feature another stable ``twin'' or ``third family'' branch
\cite{Haensel:1983,Glendenning:2000,Schertler:2000}.
Alternatively, the phase transition could occur at a 
low critical pressure:
this will be the topic of Sec.~\ref{sec:strange-dwarfs}.

%%%%%%%%%%%%%%%%%%%%%%%%%%%%%%%%%%%%%%%%%%%%%%%

\subsection{The Sturm-Liouville spectrum}
\label{sec:SL}

The BTM stability criteria are a convenient heuristic, but the fundamental
criterion for stability is based on computing the spectrum of radial
oscillations of the star. The radial oscillations
are described \cite{Chandrasekhar} by the time-dependent displacement
\beq
\de r_n(r,t) = \dfrac{e^{\nu(r)}}{r^2}u_n(r) \, e^{i \om_n t}
\label{eq:eigfunc}
\eeq
where $n$ is an index for the spectrum of modes 
and $u_n(r)$ is a solution with eigenvalue $\om_n^2$ to the 
Sturm-Liouville eigenvalue problem
\beq
\frac{\mathop{d}}{\mathop{dr}}\left(\Pi(r) \frac{\mathop{du_n}}{\mathop{dr}}\right) + \bigl( Q(r) + \om_n^2 W(r) \bigr)\, u_n(r) = 0,
\eeq
where
\begin{align*}
\Pi(r) &= \dfrac{e^{\la(r)+3\nu(r)}}{r^2}\Ga(r) P(r) \ ,\\[2ex]
Q(r) &= -4\dfrac{e^{\la(r)+3\nu(r)}}{r^3}\frac{\mathop{dP}}{\mathop{dr}} \\[1ex]
 & -8\pi \dfrac{e^{3\la(r)+3\nu(r)}}{r^2}\,P(r)\bigl(\ep(r)+P(r)\bigr) \\[1ex]
 &+ \dfrac{e^{\la(r)+3\nu(r)}}{r^2\bigl(\ep(r)+P(r)\bigr)}
  \left(\frac{\mathop{dP}}{\mathop{dr}}  \right)^2 \ ,\\[2ex]
W(r) &= \dfrac{e^{3\la(r)+\nu(r)}}{r^2}\bigl(\ep(r)+P(r)\bigr) \ ,\\[2ex]
\Ga(r) &= \dfrac{\ep(r)+P(r)}{P(r)}\dfrac{\mathop{dP}}{\mathop{d\ep}} \ .
\end{align*}

The boundary conditions for the eigenvalue problem are
\begin{align}
 u_n \propto r^3 \quad\text{at}\ r=0 \\
 \dfrac{\mathop{du_n}}{\mathop{dr}} = 0 \quad\text{at}\  r=R
\end{align}
where $R$ is the surface of the star.

The solutions to the Sturm-Liouville eigenvalue problem are a discrete set of
eigenfunctions $u_n(r)$ with eigenvalues $\om_n^2$ which are the squared
frequencies of the oscillation modes.  The eigenvalues, which are real, form a lower-bounded 
infinite sequence $\om_0^2 < \om_1^2 < \om_2^2 < \cdots $.  
For the $n$th mode, if $\om_n^2 > 0$, the
frequency is real and the mode is stable and oscillatory. However, if
$\om_n^2 < 0$ then the frequency is purely imaginary and the mode 
is unstable and exponentially grows or decays. 

To determine the overall stability of the star, it is sufficient to look just at the lowest eigenvalue, $\om_0^2$.  If $\om_0^2 > 0$, then all $\om_n^2>0$ and the star is stable.  If $\om_0^2 < 0$, then there is (at least) one unstable mode and the star is unstable \cite{shapiro}.

%%%%%%%%%%%%%%%%%%%%%%%%%%%%%%%%%%%%%%%%%%%%%%%

\section{First-order transitions and strange dwarfs}
\label{sec:strange-dwarfs}

In Refs.~\cite{GKW_PRL,GKW_long_paper}, Glendenning and collaborators
claimed that the BTM mass-radius criteria described in Sec.~\ref{sec:BTM}
were not valid for equations of state that contained a sharp first-order
transition from a low-density gas of ordinary matter
with degenerate electrons to a high-density
phase which in their case was strange quark matter.
The mass-radius curve for their
equation of state was similar to Fig.~\ref{fig:schematic_mass_radius} 
except that they claimed that the interval of the mass radius curve
from $c$ to $d$ was stable, and constituted a new family of stars,
``strange dwarfs''.

\begin{figure}
\includegraphics[width=\hsize]{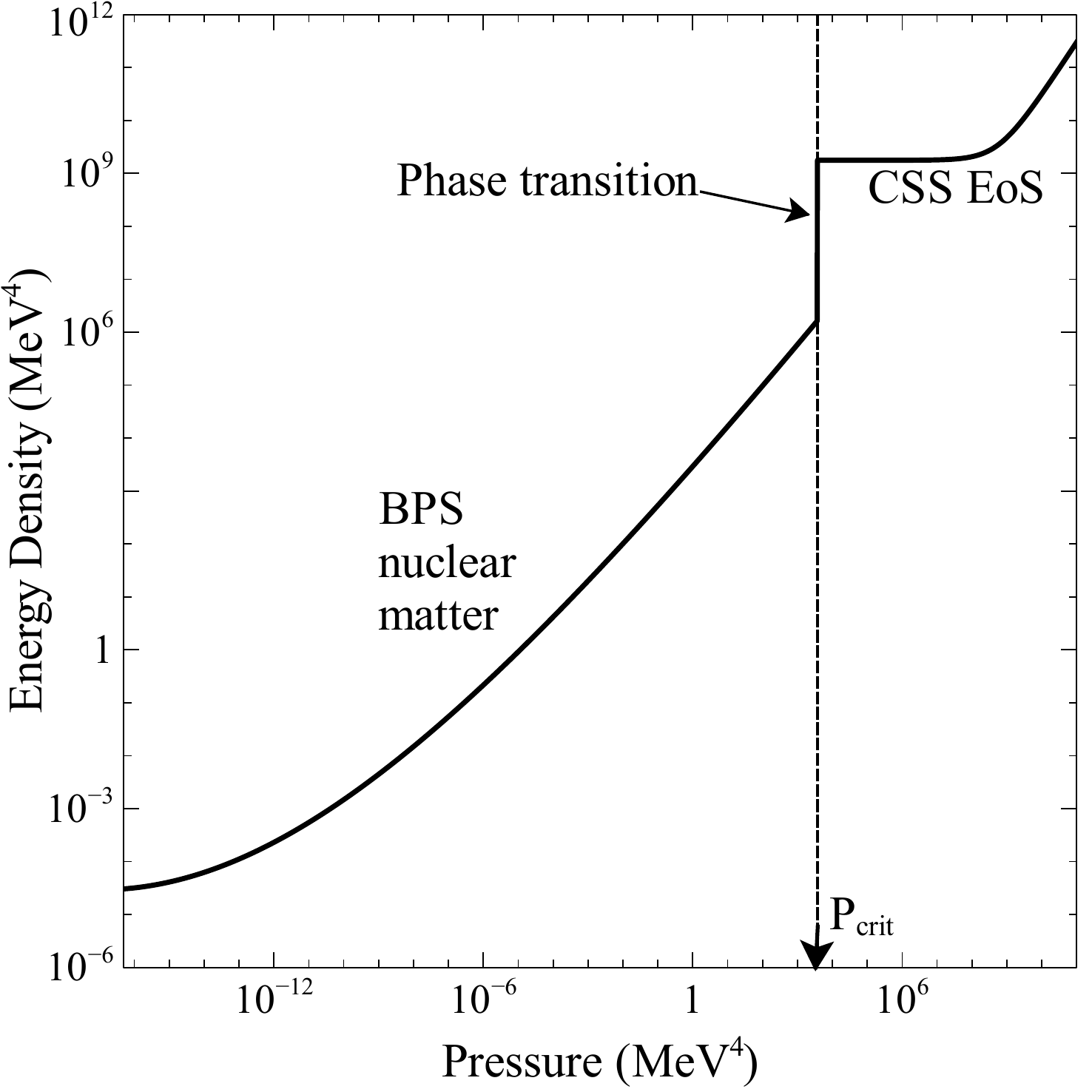}
\caption{Equation of state \eqn{eq:eos} with a sharp first-order phase transition at neutron drip pressure from BPS matter to an ultra high density phase that could, for example,
be strange quark matter.
}
\label{fig:discontinuous_eos}
\end{figure}

To study this claim, we use an equation of state similar to the
one proposed in Refs.~\cite{GKW_PRL,GKW_long_paper},
\beq
\ep(P) =  \begin{cases} 
      \ep_{\text{BPS}}(P) & P\leqslant P_{\text{crit}} \\
      kP+4B &  P> P_{\text{crit}} \ . \\
   \end{cases}
\label{eq:eos}
\eeq
This equation of state is plotted in Fig.~\ref{fig:discontinuous_eos}.
At low pressure it is the Baym-Pethick-Sutherland (BPS) 
equation of state for degenerate matter \cite{BPS}.
At a critical pressure $P_{\text{crit}}$ there is a sharp 
first-order transition
with a very large (by a factor of order $10^3$) 
discontinuity in the energy density
to a phase that is modeled
using a constant-sound-speed (CSS) equation of state
\cite{Zdunik:2012dj,Chamel:2012ea,Alford:2013aca}. This could correspond to
some exotic phase such as strange quark matter.

Glendenning~\textit{et al.}~chose $P_{\text{crit}} = P_{\text{drip}}=\SI{3742}{\MeV^{4}}$, which
is the pressure in the BPS equation of state corresponding to neutron drip
density, $\ep_{\text{drip}} = \SI{4e11}{g. cm^{-3}}=\SI{1.6e6}{MeV^{4}}.$ For
their physical model this is the highest possible transition pressure, since
if it were any larger neutrons would drip out of the crust and be attracted
into the strange matter core\cite{GW_eos}.  Additionally, they chose $k = 3$
and $B^{1/4} = \SI{145}{MeV}$.

To study the stability of stellar configurations made of matter obeying this
equation of state, we regulated the phase transition between nuclear matter
and strange quark matter, smoothing the first-order jump into a 
crossover with width $\de P$. 
This allows us to solve the Sturm-Liouville eigenvalue problem using 
standard numerical tools. These would fail if one tried to directly tackle the
discontinuous equation of state, but by sending $\de P$ to small enough 
values we can see what the limiting behavior in the discontinuous case will be.

Our equation of state took the form

\begin{align}
\ep(P) = &\frac{1}{2}\left( 1 - \tanh{\left(  \frac{P-P_{\text{crit}}}{\delta P} \right)}  \right)   \ep_{\text{BPS}}(P)\notag \\  
+ &\frac{1}{2} \left( 1+\tanh{\left(  \frac{P-P_{\text{crit}}}{\delta P} \right)} \right) \left(kP+4B \right) 
\label{eq:reg_eos}
\end{align}

We studied the solutions of the TOV equation and their stability,
according to both the BTM mass-radius criteria and explicit numerical
solution of the Sturm-Liouville eigenvalue problem, in the limit
$\de P\to 0$ where the crossover became very rapid, approximating
a discontinuity.
For the low pressure region we fitted the BPS tabulated data to a continuous 
function.

\begin{figure}
\includegraphics[width=\hsize]{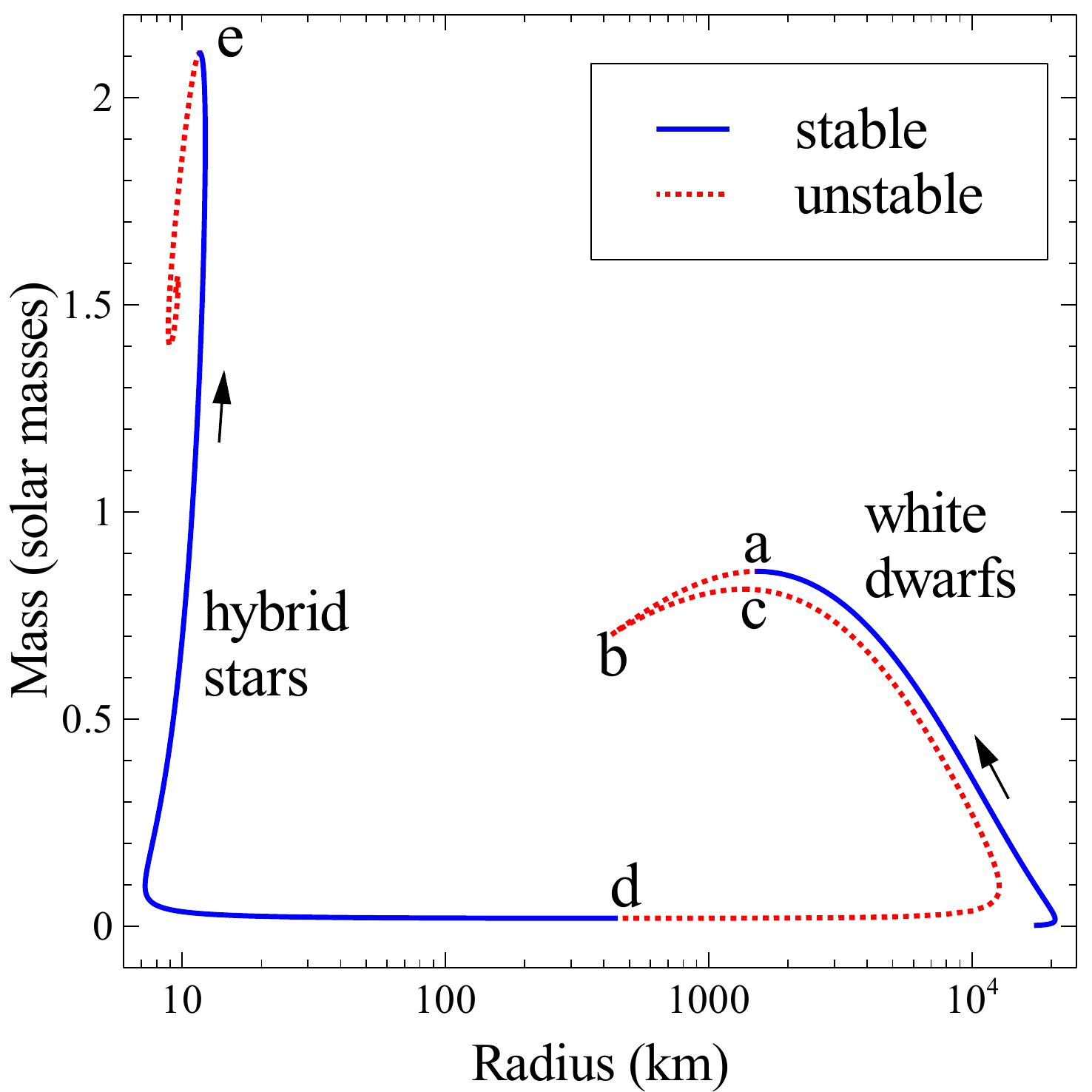}
\caption{Mass-radius plot for hybrid stars with the equation of state
given in Eq.~\eqn{eq:eos}. Solid (dashed) lines are for configurations
that are stable (unstable) according the BTM mass-radius criteria.  
The arrows indicate the direction of increasing central pressure.}
\label{fig:massradius}
\end{figure}

\section{Results and conclusions}
\label{sec:results}

In Fig.~\ref{fig:massradius} we plot the mass radius curve for solutions to
the TOV equation for the regulated equation of state ~\ref{eq:reg_eos}
with $\delta P = \SI{100}{MeV^{4}}$.

This mass-radius curve looks qualitatively like the schematic shown in Fig.~\ref{fig:schematic_mass_radius}. The extremum labelled $b$ occurs where the
central pressure in the star reaches the critical pressure $P_{\rm crit}$.
This means that all stars with central pressure below the value at $b$ 
(i.e.~on the curve from $b$ to $a$ and further along the white dwarf branch)
have only conventional degenerate-electron
matter: the center of the star is not yet dense enough to 
contain any of the high-density phase.
For central pressures larger than at $b$, the star contains a core of
the high-density phase. On the curve from $b$ through $c$ and on to $d$
the core is small relative to the surrounding crust of nuclear matter.
Glendenning \textit{et al.}, using a model where the high-density phase is
strange quark matter, called these stars ``strange dwarfs''.
At higher central pressures, on the curve from $d$ to $e$, the core becomes
large and its gravitational attraction squeezes the crust down to a thin
layer: In Glendenning \textit{et al.}'s model these are strange quark stars
with a thin nuclear crust.

If there is a sharp first-order transition in the equation of state
then point $b$ is a cusp in the $M(R)$ relation. For finite but very
small transition width $\de P\lesssim 1\,\MeV^4$ the cusp becomes a
minimum at which, according to the BTM criteria,
the second-lowest mode goes from stable to unstable
as central pressure increases.
In our calculation we use values of $\de P$ in the range $10$ to
$100\,\MeV^4$, in which case the mass radius relation develops a
more complicated structure at $b$ which may have multiple extrema
as the curve spirals and then ``uncoils'' again. 
This structure occurs in a very small range of masses and radii near $b$,
and is invisible on the scales shown in Fig.~\ref{fig:massradius}.
The details of this structure depend on the exact profile of the regulated 
transition, but, as we will see,
(i) the lowest eigenmode remains negative so all these configurations are unstable; (ii) as central pressure increases through $b$, the net outcome
is that the second-lowest mode goes from stable to unstable;
(iii) this behavior is not relevant to the stability of strange dwarfs,
which lie between $c$ and $d$ on the mass-radius curve.

According to the BTM mass-radius criteria (Sec.~\ref{sec:BTM}), only the
portions of the mass-radius curve denoted by a solid line in Fig.~\ref{fig:massradius} are stable. However, Glendenning \textit{et al.} claim that the
portion between extrema $c$ and $d$, corresponding to the strange dwarfs, 
is also stable. To test this we solved the Sturm-Liouville 
eigenvalue problem for stars with a range of central pressures.

\begin{figure}
\includegraphics[width=\hsize]{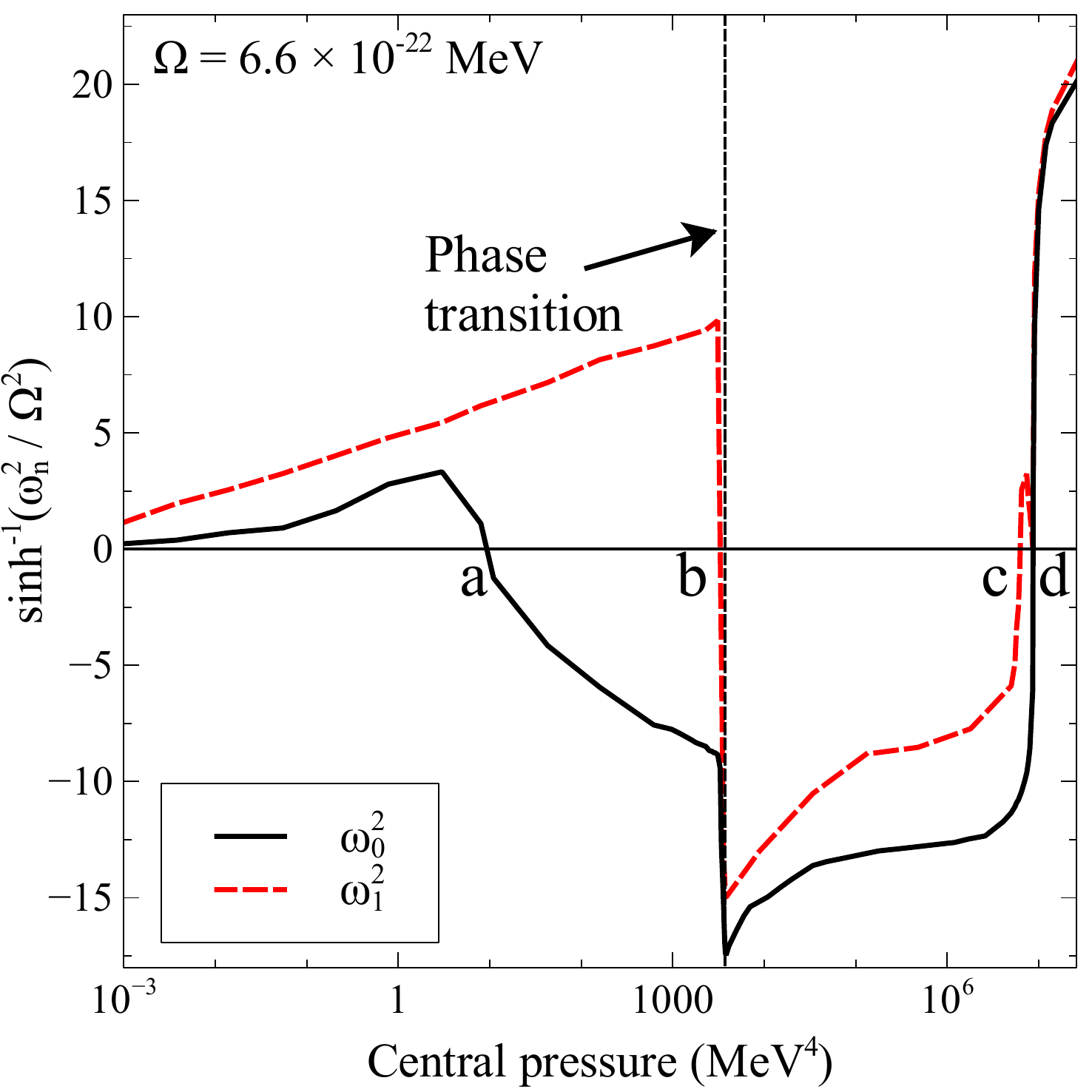}
\caption{The squared frequencies of the two
lowest radial oscillations for the TOV solutions plotted in
Fig.~\ref{fig:massradius}. For stellar configurations with 
central pressures between $a$ and $d$ we find that $\omega_0^2 < 0$ so
these configurations are unstable. This agrees with
the BTM criteria \eqn{eq:BTM-criteria}.
}
\label{fig:SL1}
\end{figure}

\begin{figure}
\includegraphics[width=\hsize]{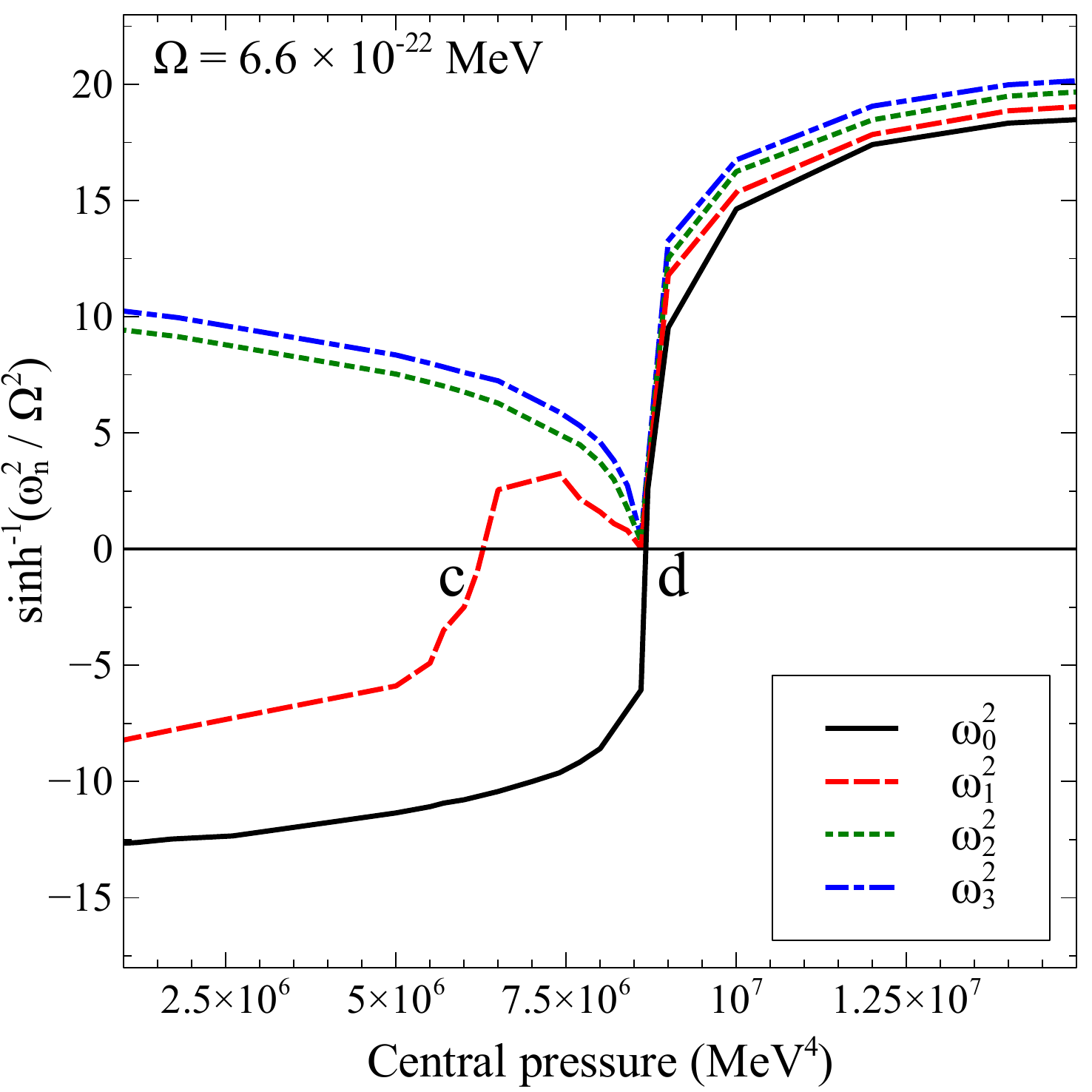}
\caption{
The squared frequencies of the four
lowest oscillation modes, as in Fig.~\ref{fig:SL1}, but magnified to
more clearly show the range of central
pressures from $c$ to $d$ where strange dwarfs were hypothesized to occur.  Between $c$ and $d$, $\omega_0^2 < 0$ and thus strange dwarfs are unstable.
}
\label{fig:SL2}
\end{figure}

\begin{figure}
\includegraphics[width=\hsize]{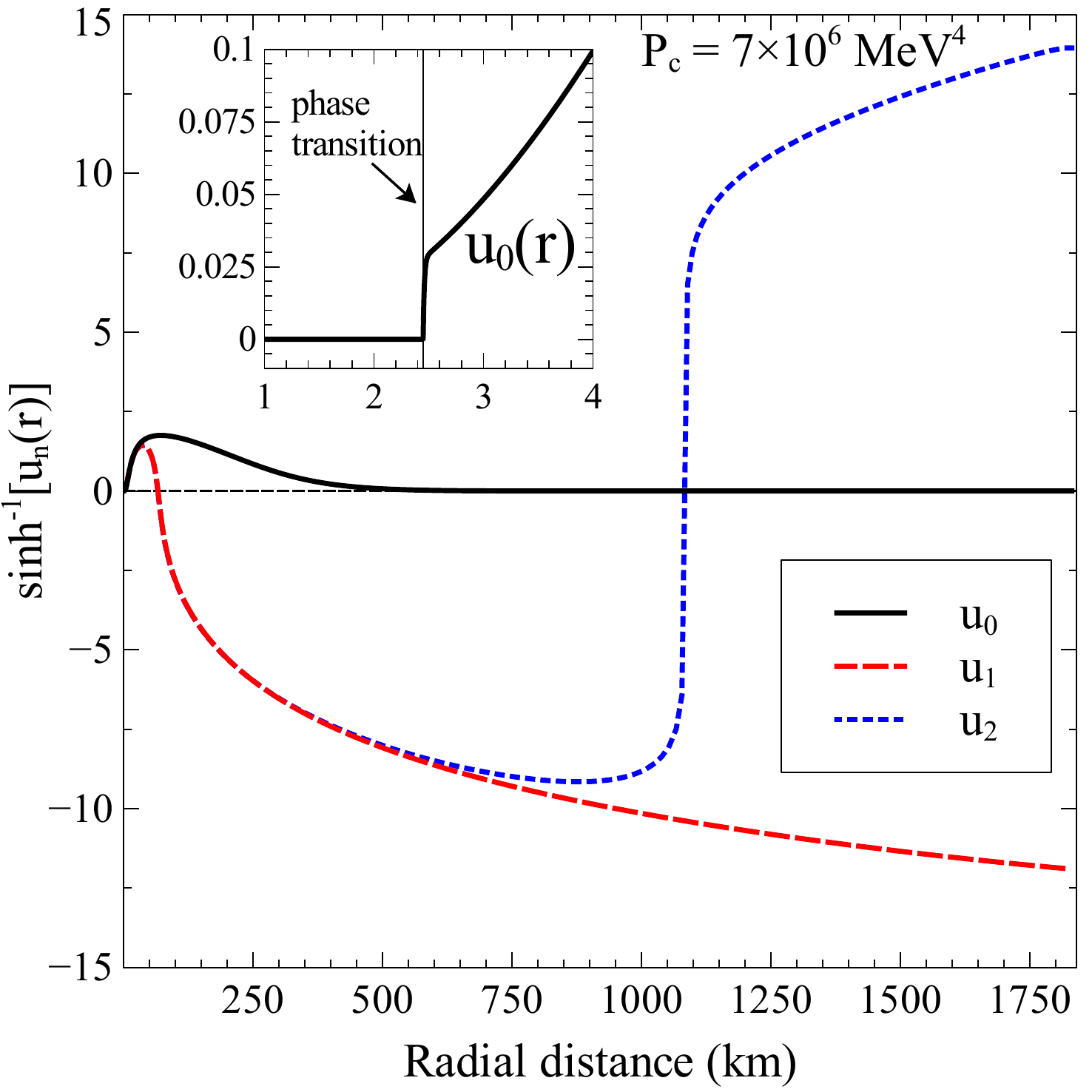}
\caption{The $n=0, 1, \text{and } 2$ eigenfunctions $u_{n}(r)$ (see Eq.~\ref{eq:eigfunc}) for a TOV solution with central pressure \SI{7e6}{\MeV^{4}}, which lies between points $c$ and $d$ in Fig.~\ref{fig:massradius}.  At this central pressure, the phase transition is located at $r=\SI{2.4}{\km}$.  As expected for Sturm-Liouville eigenfunctions, $u_n(r)$ has $n$ nodes. The arcsinh scale exaggerates the sharpness of the zero-crossing of the $n=2$ eigenfunction.}
\label{fig:eigfunc}
\end{figure}

\begin{figure}
\includegraphics[width=\hsize]{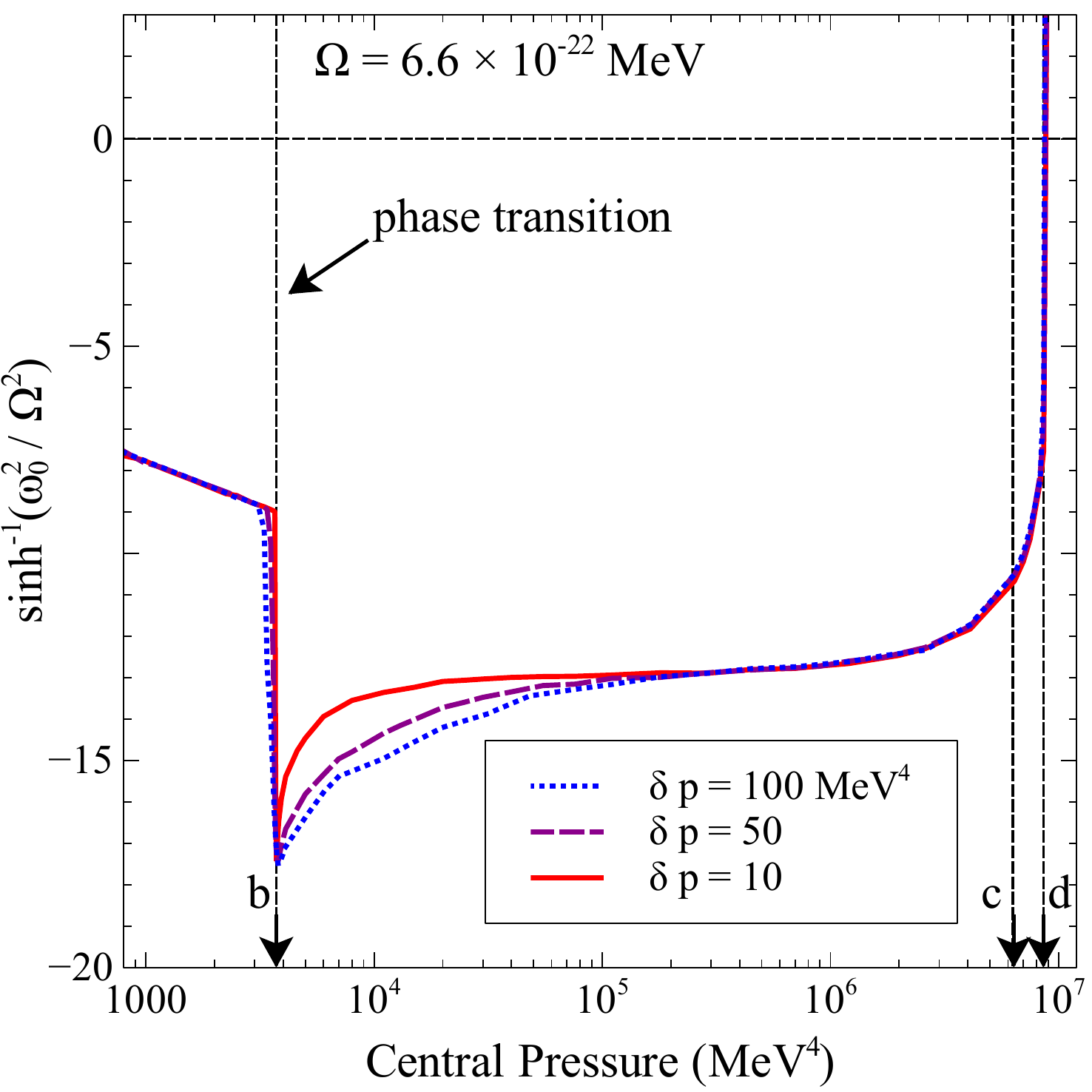}
\caption{Dependence of the lowest squared frequency on the regulator $\de P$ that gives the phase transition a non-zero width.
In the limit $\delta P \rightarrow 0$, the lowest eigenvalue $\omega_0^2$ remains negative for all central pressures between $a$ and $d$.}
\label{fig:SL3}
\end{figure}

Our results are displayed in 
Figs.~\ref{fig:SL1},\ref{fig:SL2},\ref{fig:eigfunc},\ref{fig:SL3}, where
we have used an arcsinh scale on the y axes. This has the wide dynamic range of
a log scale while also including zero and negative values.

In Fig.~\ref{fig:SL1} we show the two lowest
eigenvalues, $\om_0^2$ and $\om_1^2$, as a function of central pressure.
We see that these behave in a way that is consistent with the
BTM mass-radius criteria: at extremum $a$ the lowest mode becomes unstable
($\om_0^2$ goes below zero). At $b$ the next-to-lowest mode also
becomes unstable, and then at extremum $c$ it becomes stable again.
The lowest mode remains unstable until we reach extremum $d$, at which point
it becomes stable again.

In Fig.~\ref{fig:SL2} we zoom in on the range of central pressures between
extrema $c$ and $d$, where Glendenning \textit{et al.} claim that there are
stable strange dwarf configurations. 
In this range the lowest radial eigenmode 
remains negative, indicating that these solutions to the TOV equation
are unstable. Comparing with Fig.~2 in Ref.~\cite{GKW_PRL}, it seems likely
that Glendenning \textit{et al.} mistook the second-lowest eigenmode for the
lowest one, giving them the impression that these configurations were stable.

To check that we have found the lowest eigenmode we show
in Fig.~\ref{fig:eigfunc} the eigenfunctions of a configuration with central pressure of $\SI{7e6}{\MeV^{4}}$, which lies between extrema $c$ and $d$.  
The $u_0$ eigenmode has no nodes, indicating that it is indeed the
lowest mode. In general the $n$-th eigenfunction has $n$ nodes, 
as expected for solutions to a Sturm-Liouville problem.

The eigenvalue spectra shown in Figs.~\ref{fig:SL1} and \ref{fig:SL2} were calculated for a regulator width $\delta P = \SI{100}{MeV^{4}}$.  To show that the results carry over to the discontinuous limit, we show in Fig.~\ref{fig:SL3} the dependence of $\om_0^2$ on central pressure for several different regulator widths.  The spectrum shows some dependence on the regulator width when
$P_{\rm cent}$ is close to $P_{\rm crit}$ (near $b$ on the $M(R)$ curve), which is expected since this is where a tiny core of the high density phase first appears in the star.

However, the lowest eigenvalue remains negative as the transition becomes 
sharper ($\de P\to 0$), and in the strange dwarf region (between $c$ and $d$)
there is very little sensitivity to the regulator.

We conclude that the BTM mass-radius criteria for the stability of
stars are valid in the presence of an arbitrarily sharp jump in the
energy density as a function of pressure. Our results imply that the 
strange dwarfs proposed in Refs.~\cite{GKW_PRL,GKW_long_paper} are not
stable.

\begin{acknowledgments}
We thank Fridolin Weber for discussions.
This material is based upon work partly
supported by the U.S. Department of Energy, 
Office of Science, Office of Nuclear Physics under Award Number
\#DE-FG02-05ER41375.
\end{acknowledgments}

%\clearpage

%%%%%%%%%%%%%%%%%%%%%%%%%%%%%%%%%%%%%%

\bibliographystyle{JHEP}
\bibliography{compact_star_stability}{}

\end{document}